\documentclass[aps,prl,twocolumn,showpacs,superscriptaddress]{revtex4-1}
\usepackage{graphicx, epsfig, amssymb, amsmath, lineno}

\begin{document}

\title{Local Wettability Reversal during Steady-State Two-Phase Flow in 
Porous Media}

\author{Santanu Sinha}
\email{Santanu.Sinha@ntnu.no}
\author{Morten Gr{\o}va}
\author{Torgeir Bryge {\O}deg{\aa}rden}
\affiliation{Department of Physics, Norwegian University of Science and
  Technology, N-7491 Trondheim, Norway}
\author{Erik Skjetne}
\affiliation{Statoil, Arkitekt Ebbellsvei 10, N-7005 Trondheim, Norway}
\author{Alex Hansen}
\affiliation{Department of Physics, Norwegian University of Science and
  Technology, N-7491 Trondheim, Norway}

\date{\today}

\begin{abstract}
We study the effect of local wettability reversal on remobilizing
immobile fluid clusters in steady-state two-phase flow in porous
media. We consider a two dimensional network model for porous medium
and introduce a wettability alteration mechanism. A qualitative change
in the steady state flow patterns, destabilizing the percolating and
trapped clusters, is observed as the system wettability is
varied. When capillary forces are strong a finite wettability
alteration is necessary to move the system from single-phase to
two-phase flow regime. For the case of both phases being mobile we
find a linear relationship between fractional flow and wettability
alteration.

\end{abstract}

\pacs{47.56.+r,47.61.Jd}

\maketitle

In a time of dwindling oil reserves, the fact that some 20 to 60
percent of the oil remains unrecovered at the end of oil production
constitutes a challenge of increasing importance \cite{r05}. The
reason for this loss is the formation of oil clusters embedded in
water and held in place by capillary forces, which in turn are
controlled by the wetting properties of the reservoir fluids with
respect to the matrix rock. Therefore, the wetting properties of crude
oil, brine and rock systems is a central research topic in the
petroleum industry within the field of Enhanced Oil Recovery
(EOR). EOR covers recovery methods other than pressure depletion
(typical primary recovery) and water or gas injection (typical
secondary recovery).

Formation wettability was reviewed recently in \cite{a07}. Reservoir
rocks do not have uniform wetting properties. When a reservoir
fluctuates spatially between strongly oil-wet and strongly water-wet
it is called fractionally wet \cite{b98}. Sandstone is strongly
water-wet before oil migrates from a source rock into the
reservoir. After oil enters a pore it forms water films \cite{h91a,
  i11} and a wettability alteration of the rock is believed to take
place which can be irreversible due to direct adsorption of
asphaltenes \cite{k93, k98} or can be reversible due to other surface
active molecules present in the crude oil, such as caboxyl acids.

Both laboratory experiments and field tests have shown that deviation
from strongly oil-wet to water-wet or neutral-wet conditions
significantly increase oil recovery efficiency \cite{tht99}.
Wettability of reservoir rocks can be altered by changing the brine
composition, e.g., lowering salinity \cite{tm97}, adding water-soluble
surfactants \cite{sa00} or even by adding oil-soluble organic acids or
base \cite{tht99}. An increase in temperature also increases water
wetness of the reservoir \cite{stk06}. Correlations have been shown
with wetting behavior to the electrostatic forces between the mineral
and oil surfaces \cite{btm89}. Increasing the salinity level of the
pore water makes the wetting transition shift in pH. The oil industry
therefore paying a great attention towards low-salinity water
flooding.

So far, there is no consensus on the dominating microscopic mechanisms
behind the transport of the agent that mediates the change in the pore
wettability. The result of it, however, will be that there are
correlations between the positions of the altered pores. In light of
the massive uncertainty connected to the transport mechanisms, we
simplify the problem assuming no correlation between positions of the
altered pores. The precise question we pose in this paper is the
following: Will the reversal of the wetting properties of a percentage
of the pores lead to a significant change in the fractional flow
patterns?  We will see here that even without wettability switch
correlations there can be a permanent change in fractional flow when
the switched fraction of pores is large enough. How spatial
correlations between the altered pores further affect the flow is a
matter for future investigations.

%Many low-salinity flooding mechanisms have been suggested, but there
%is no consensus on what is the dominating microscopic mechanism.

%Wettability of reservoir rocks are altered by changing the brine
%composition, e.g., lowering salinity \cite{tm97}, adding water-soluble
%surfactants \cite{sa00} or even by adding oil-soluble organic acids or
%base \cite{tht99}.

In order to shed light on how local wettability alterations may
remobilize stuck fluid clusters and to look for the consequences of
this on the flow properties, we study here the effect of such
alterations on a two-dimensional pore scale transport network model
\cite{amhb98, kah02, rh06}. The mechanism for local wettability
alteration is introduced in the model by changing the wetting contact
angles of the menisci at pore necks. To study steady-state flow
reflecting the situation deep inside the reservoir, we implement
biperiodic boundary conditions. The network is represented by a square
lattice of cylindrical tubes, tilted $45^\circ$ with respect to the
imposed pressure gradient as well as the overall flow
direction. Disorder is incorporated in the system by assigning the
radius ($r$) of each tube randomly from a flat distribution in the
range $[0.1l, 0.4l]$, where $l$ is the length of the tube. The network
is filled with two immiscible fluids, named as water and oil, which
flow inside the tubes (links). With respect to a pore, one fluid is
less wetting while the other is more --- hence there is a contact
angle $\theta$ associated with each meniscus. The tubes are
cylindrical with respect to permeability, but they are considered
hour-glass shaped with respect to the capillary pressure $p_c$. At
position $x$, $p_c$ is obtained from a modified form of the
Young-Laplace law \cite{amhb98, d92},
\begin{equation}
\displaystyle
p_c = \frac{2\gamma \cos \theta}{r}\big[1 - \cos\frac{2\pi x}{l}\big]\;,
\label{pc}
\end{equation}
where $\gamma$ is the interfacial tension between the fluids. We
define the wettability of the system by $\omega =
N_\textnormal{oil}/N_\textnormal{total}$, where $N_\textnormal{oil}$
is the number of oil-wet links among the total $N_\textnormal{total}$
links in the system. Thus $\omega = 0$ and $\omega = 1$ represent pure
water-wet and pure oil-wet systems respectively. A fractional value of
$\omega$ represents a fractionally wet system. When the wettability of
a pore is altered, the direction of capillary forces at the menisci
inside that pore changes. We therefore change the wettability of a
pore in the model by altering the sign of $p_c$. In this work we
consider a mixture of perfectly oil-wet and perfectly water-wet
conditions, i.e., $\theta$ is either $0^\circ$ or $180^\circ$.

The flow is driven by setting up an external global pressure drop. The
local flow rate $q$ in a tube with a pressure difference $\Delta p$
between the two ends of that tube follows the Washburn equation of
capillary flow \cite{w21, d92},
\begin{equation}
\displaystyle
q = -\frac{a k}{\mu l}\big(\Delta p - \Sigma p_c\big)\;,
\label{wb}
\end{equation}
where $k$ is the permeability. For cylindrical tubes, $k=r^2/8$ which
is known from Hagen-Poiseuille flow. Here $a$ is the cross-sectional
area of the tube and $\mu$ is the volume average of the viscosities of
the phases present inside the tube. The sum over $p_c$ runs over all
menisci within the tube. The net flow at a node must be zero in a
network. Eq.\ \ref{wb} then gives a large number of linear equations
connecting the pressure variables at the nodes. The pressure field is
obtained by solving the linear equations \cite{bh88} and the flow
field then follows from eq.\ \ref{wb}. The system is forward time
integrated by one step, using the explicit Euler scheme. Inside a tube
all menisci move with a front speed determined from the local flow
rate $q$, but when a meniscus reaches the end of a tube new menisci
are formed in neighboring tubes maintaining the volume conservation. A
maximum number of six menisci is considered here in a given
tube. Further details of how the menisci are moved can be found in
\cite{kah02}, though they are not necessary to understand the results
of wettability alteration here.

The flow is controlled by the ratio between capillary and viscous
forces at the pore level and quantified by capillary number
$\textnormal{Ca}=\mu Q_\textnormal{tot}/(\gamma \Sigma)$ where
$\Sigma$ is the cross-sectional area of the network and
$Q_\textnormal{tot}$ is the total flow rate. Two different capillary
numbers $\textnormal{Ca} = 1.0\times 10^{-2}$ and $\textnormal{Ca} =
1.0\times 10^{-3}$ are considered here. The ratio between the
viscosities of the two fluids, ${\rm M} = \mu_{\rm o} /\mu_{\rm w}$ is
another dimensionless parameter to control the flow, where $\mu_{\rm
  o}$ and $\mu_{\rm w}$ are the viscosities of oil and water
respectively. Here we restrain our discussion to viscosity matched
fluids which gives $M=1$. Simulation is done considering a network of
$40\times 40$ links. This is sufficiently large to be in the
asymptotic limit for the range of parameters \cite{kah02}. An average
over $10$ different samples has been taken for each simulation.

\begin{figure}[t]
\centerline{
\psfig{file=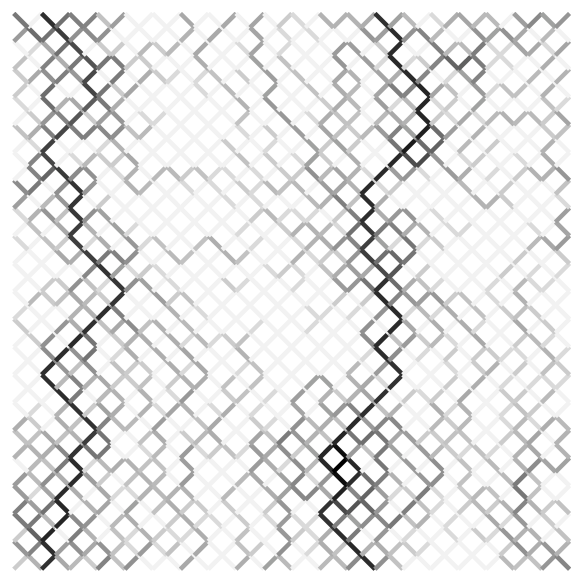,height=.18\textwidth}
\psfig{file=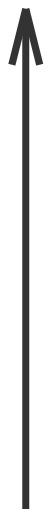,height=.18\textwidth}
\hfill
\psfig{file=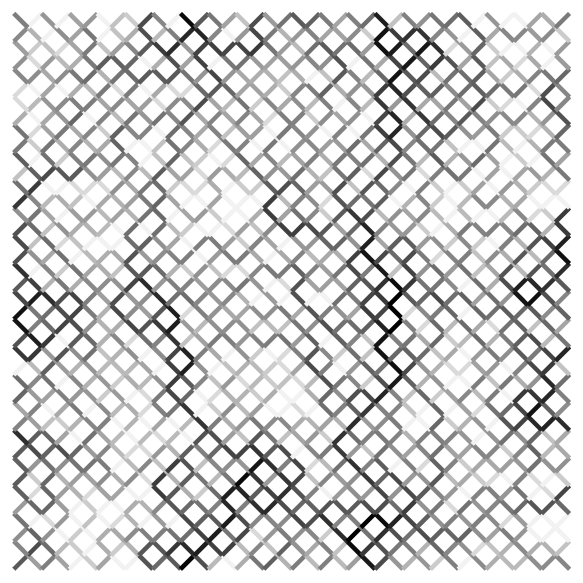,height=.18\textwidth}
\hfill
\psfig{file=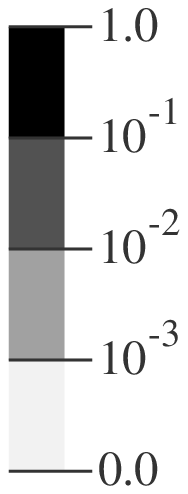,height=.18\textwidth}
}
\centerline{\hfill (a) \hfill \hfill \hfill (b) \hfill \hfill}
\caption{\label{snap}Distribution of normalized local flow rates ($q$)
  (a) before and (b) after the wettability alteration at
  $\textnormal{Ca} = 1.0\times 10^{-3}$ and $S=0.6$. The long arrow
  indicates the overall flow direction. The highest to lowest flow
  rates are indicated by darkest to lightest gray as shown by the
  scale. The steady state flow before wettability alteration is
  dominated inside only a few percolating paths as seen in (a). In
  (b), wettability of $50\%$ tubes are altered and the percolating
  path breaks, causing the flow to be distributed over the whole
  network.}
\end{figure}

The simulation is performed under constant flow rate
$Q_\textnormal{tot}$ which sets the capillary number. Initially, the
system is pure water-wet ($\omega=0$) and filled with given oil and
water saturations which remain constant throughout the
simulation. Thus the saturation, capillary number and the wettability
are the independent variables here. As the system evolves with time,
it either starts a viscous fingering for high values of Ca or invasion
percolation process when capillary forces dominate. Due to biperiodic
boundary conditions, both drainage and imbibition take place
simultaneously, leaving wetting and non-wetting fluid clusters in the
system. On long timescales the system evolves to a steady state
characterized by the system's macroscopic properties, such as global
pressure, that becomes constant on the average. Mixing of two phases
in steady state depends on the external control parameters, both Ca
and the saturation of fluids. In Fig.\ \ref{snap}(a), the distribution
of normalized local flow rates at steady state is shown for the lower
capillary number $\textnormal{Ca} = 1.0\times 10^{-3}$. If the
difference in the saturation of the two fluids is large, one of the
phases percolates the system with trapped immobile clusters of the
other phase as can be seen in Fig.\ \ref{snap}(a). Clearly, the flow
is dominated through a few anisotropic percolating paths \cite{ss06},
leaving trapped immobile clusters in the system. In this case, we may
apply Darcy's law for the single-phase pressure $P_s$ along the system
length $L$ by,
\begin{equation}
\frac{Q}{\Sigma '} = \frac{k}{\mu} \frac{\Delta P_s}{L}\;,
\label{darcy}
\end{equation}
which will be significantly larger than that of single-phase flow due
to the smaller effective cross section $\Sigma '$. However, in the
other situation when neither phase percolates the system and both the
phases are mobile, all spanning flow paths contain interfaces
associated with capillary forces. This situation requires even larger
driving pressure due to the local dynamics of interfaces and capillary
barriers. Therefore, as soon as low resistance percolating path
appears in the system the global pressure drop decreases and the
percolating and trapped clusters remain static. This is not favorable
for oil-recovery purposes as it leaves immobile fluid in the
reservoir.

After the system evolves to a steady state we change the wettability
$\omega$ by altering the direction of capillary forces in required
number of links, chosen randomly. This initially perturbs the pressure
field, deviates the system from existing steady state and subsequently
leads it to settle into a new steady state. This again results to
break any existing percolating and trapped fluid clusters which start
flowing within a few time steps as illustrated in
Fig.\ \ref{snap}(b). If the wettability alteration is sufficiently
large the change to a non-percolating system is permanent, otherwise
new percolating cluster eventually appears.

\begin{figure}[t]
\centerline{\hfill
\psfig{file=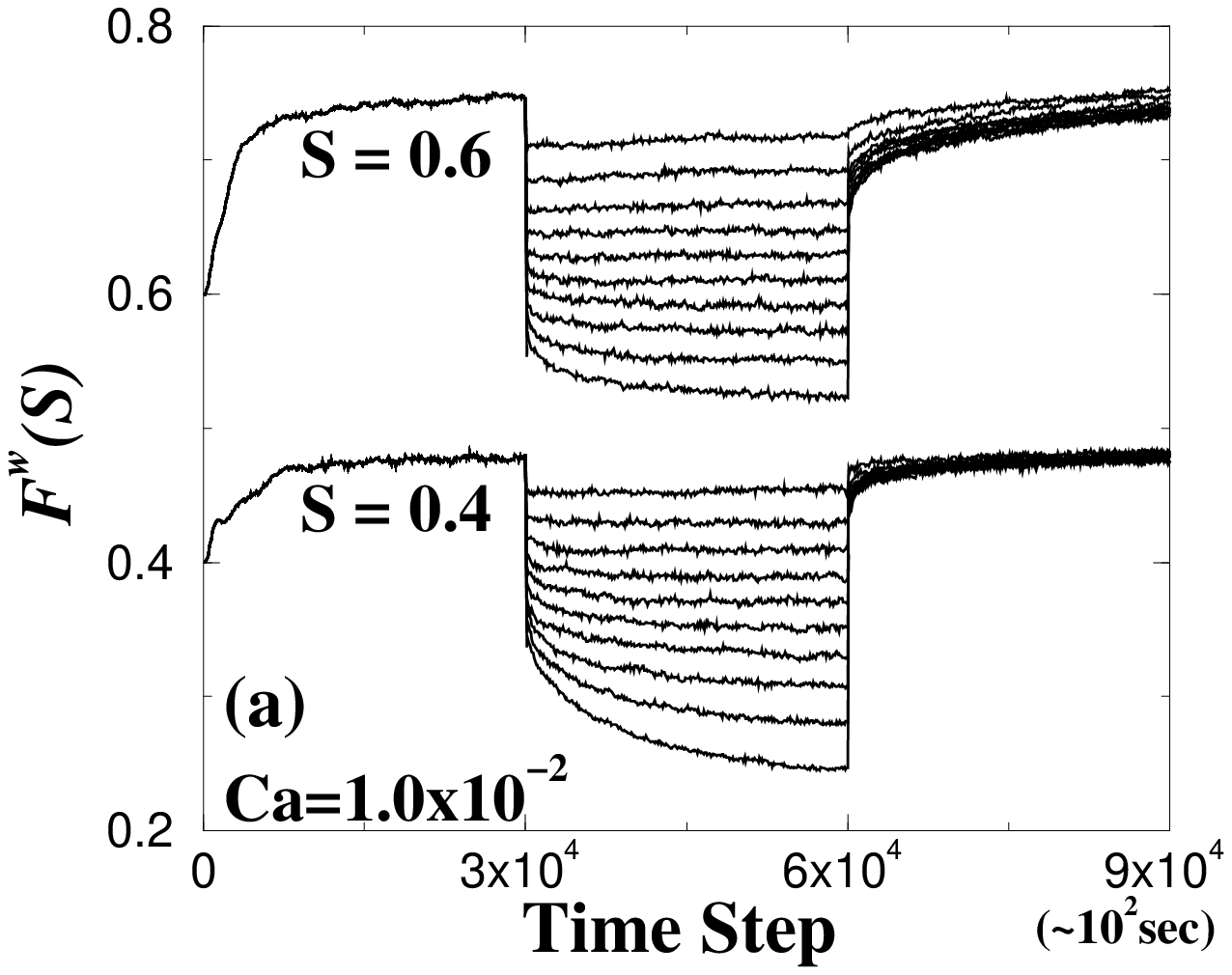,width=.22\textwidth}
\hfill
\psfig{file=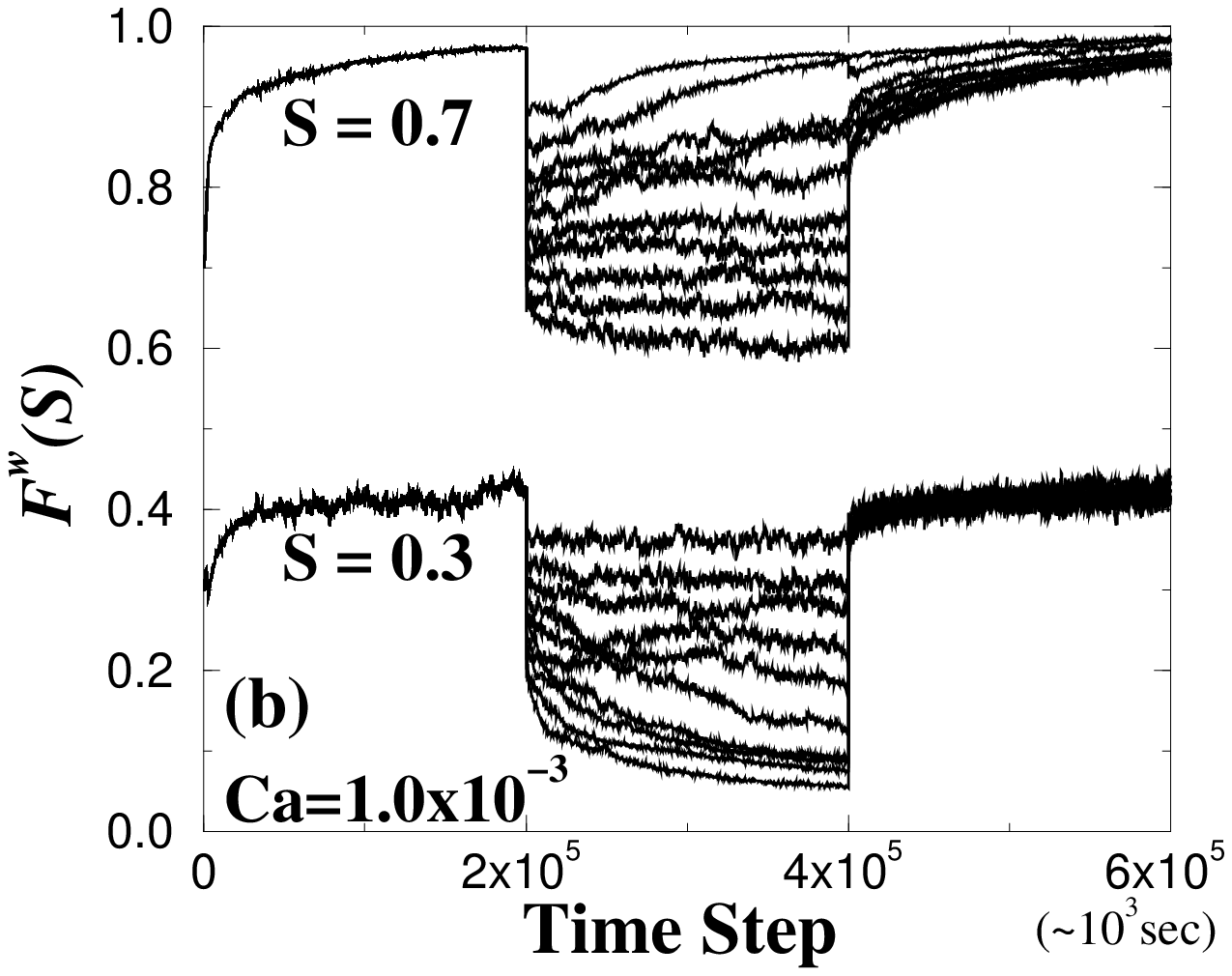,width=.22\textwidth}
\hfill}
\caption{\label{flow-t} Variation of oil fractional flow with
  time. Three different regions in each curve imply the time evolution
  in a water-wet system, then in a fractionally wet system after a
  wettability alteration $\omega$ and finally restoring the initial
  water-wet system. For each saturation $10$ different curves from top
  to bottom correspond to different values of $\omega$ from $0.1$ to
  $1.0$ in intervals of $0.1$. As $\omega$ increases a gradual change
  in fractional flow is observed.}
\end{figure}

We measure the fractional flow of oil, defined as the ratio between
the flow of oil ($Q_\textnormal{oil}$) with the total flow
($Q_\textnormal{tot}$) as $F^\omega(S)=Q_\textnormal{oil}/
Q_\textnormal{tot}$ in a system with wettability $\omega$ and oil
saturation $S$. Independent simulations have been performed for
different values of $\omega$ and $S$. In Fig.\ \ref{flow-t},
$F^\omega(S)$ is plotted as a function of time. The three regions of
the curves correspond to the initial pure water-wet ($\omega=0$)
system, flipping wettability by a nonzero value of $\omega$ and then
again flipping them back to the initial $\omega=0$ value. There is a
few things to notice. First, there is a gradual decrease in the steady
state oil fractional flow with increasing $\omega$, i.e., more
oil-wetness of the system. This is consistent with experimental
results as it is observed that changing from oil-wet to mixed and
water-wet systems, as in the case for low-salinity flooding, increases
oil production \cite{sa00,tht99}. Secondly, the alteration process is
reversible here, as in the third regime of the curves $F^\omega(S)$
returns to the initial steady state value $F^0(S)$. Third, when the
capillary forces are strong and the system is in a single-phase flow
regime, a finite amount of wettability alteration is needed to move it
towards a two-phase regime. This can be seen in Fig.\ \ref{flow-t}(b),
that for low capillary number $\textnormal{Ca} = 1.0\times 10^{-3}$
and saturation $S=0.7$, apart from an initial transient response, the
steady state fractional flow does not change for $\omega \le 0.2$.

\begin{figure}[t]
\centerline{\hfill
\psfig{file=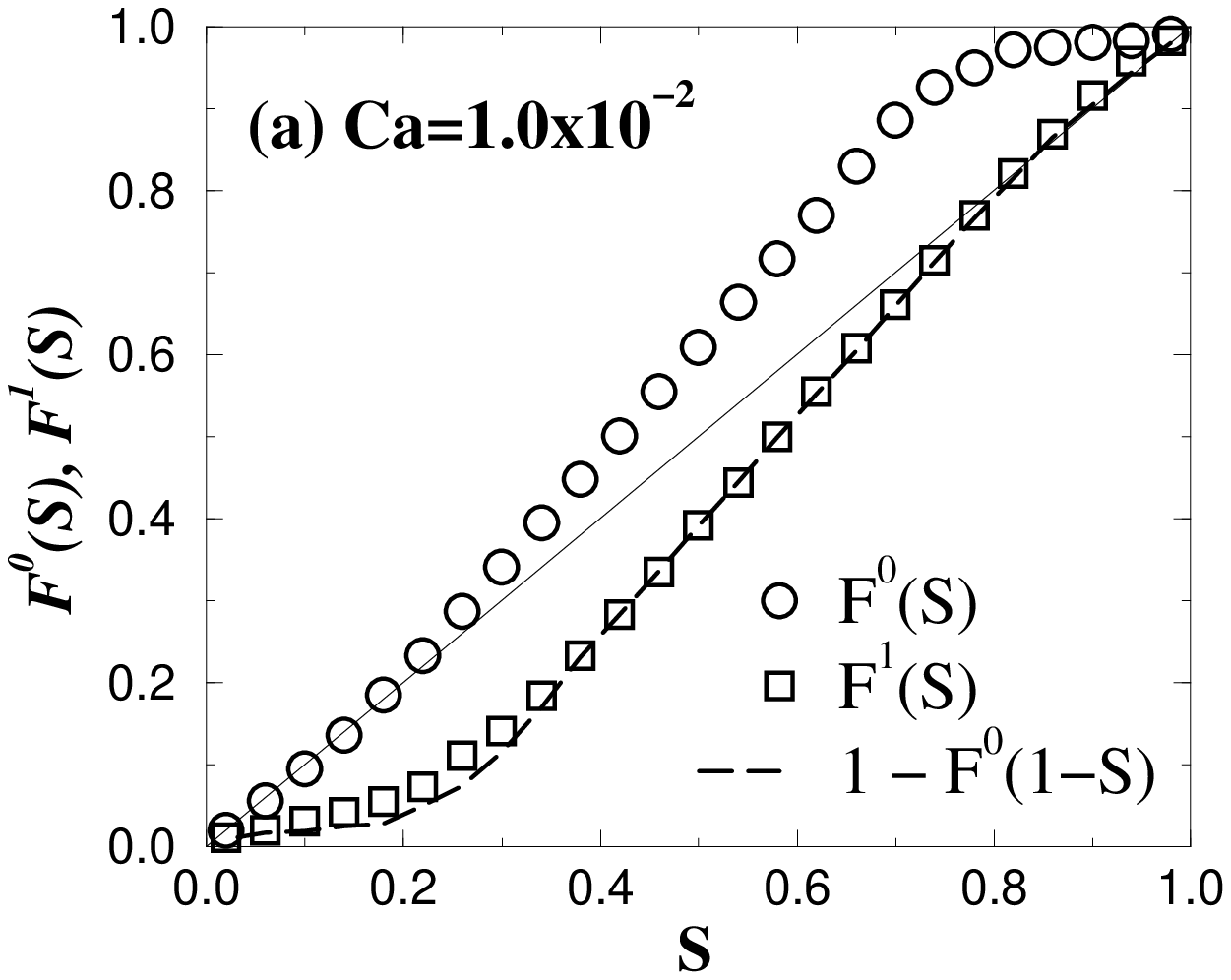,width=.22\textwidth}
\hfill
\psfig{file=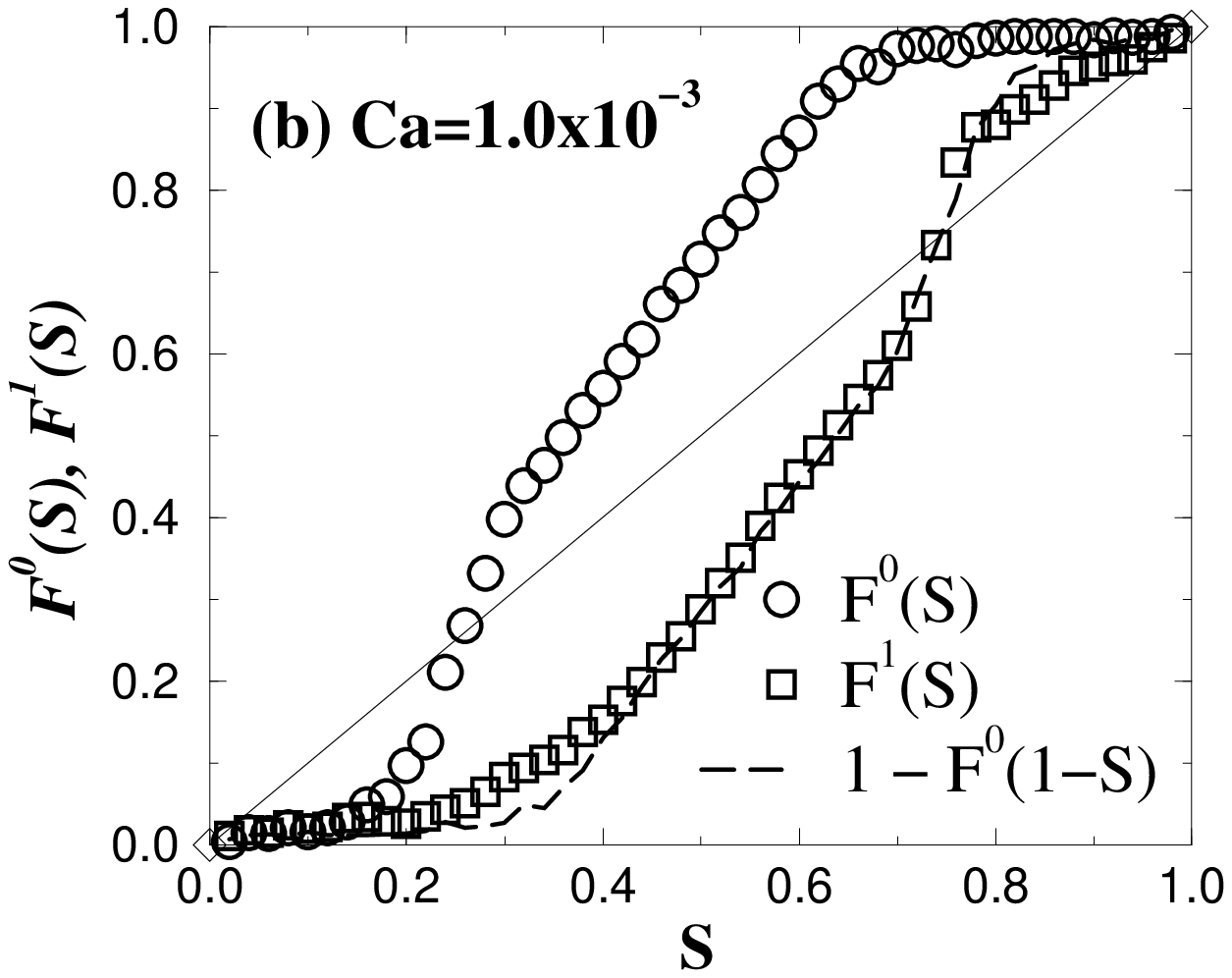,width=.22\textwidth}
\hfill}
\caption{\label{flow-s}Fractional flow as a function of saturation in
  water-wet ($\bigcirc$) and oil-wet ($\Box$) systems. $1-F^0(1-S)$,
  calculated from $F^0(S)$, is drawn by the dashed line which matches
  nicely with the $F^1(S)$ curve.}
\end{figure}

The variation of oil fractional flow with oil saturation $S$ in a pure
water-wet and a pure oil-wet systems are plotted in
Fig.\ \ref{flow-s}. In both the systems, the fractional flow does not
follow $F=S$ value represented by the solid straight line. This is due
to the presence of interfaces with capillary pressure. The curves
cross the $F=S$ line from below at some point less than $50\%$
saturation in pure water-wet and higher than $50\%$ saturation in pure
oil-wet systems. Fractional flow always higher in water-wet system
($F^0(S)$) than in the oil-wet system ($F^1(S)$). The difference
$\Delta F^1(S) = F^0(S) - F^1(S)$ depends on the saturation, it is
zero at the two end points $S=0$ and $1$. It is maximum in between,
where both the phases are mobile. The maximum value of $\Delta F$
increases with decreasing $\textnormal{Ca}$, it is $\approx 0.22$ for
$\textnormal{Ca} = 1.0\times 10^{-2}$ and $\approx 0.40$ for
$\textnormal{Ca} = 1.0\times 10^{-3}$. This due to the shape of the
curves which strongly depend on $\textnormal{Ca}$ \cite{kah02-2,kh04}.

\begin{figure}[b]
\centerline{\hfill
\psfig{file=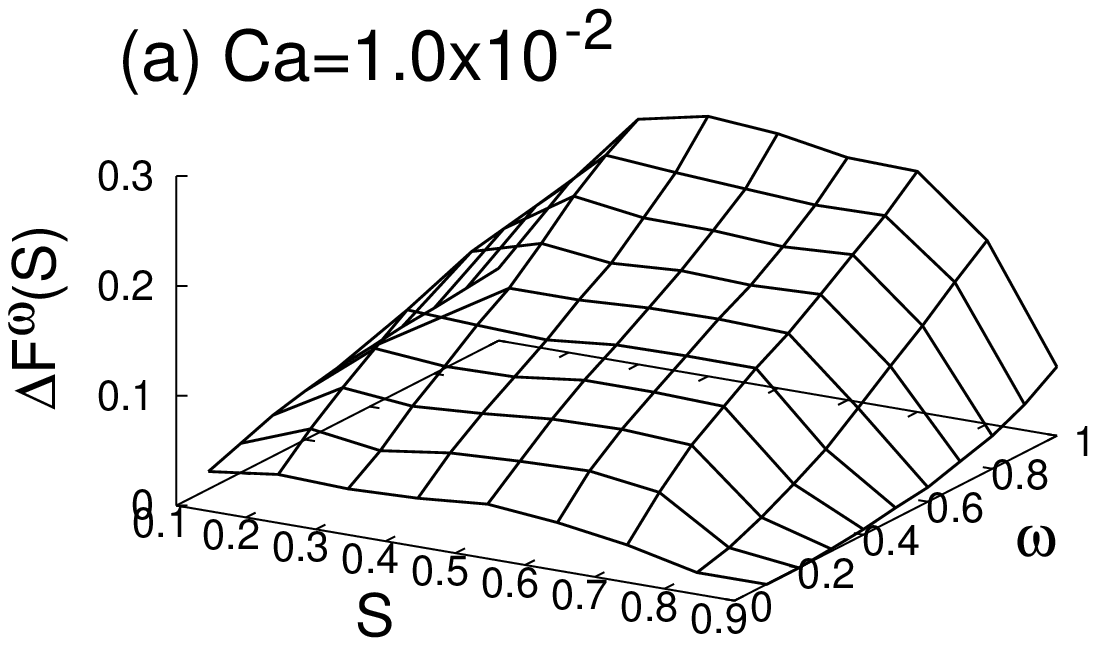,width=.22\textwidth}
\hfill
\psfig{file=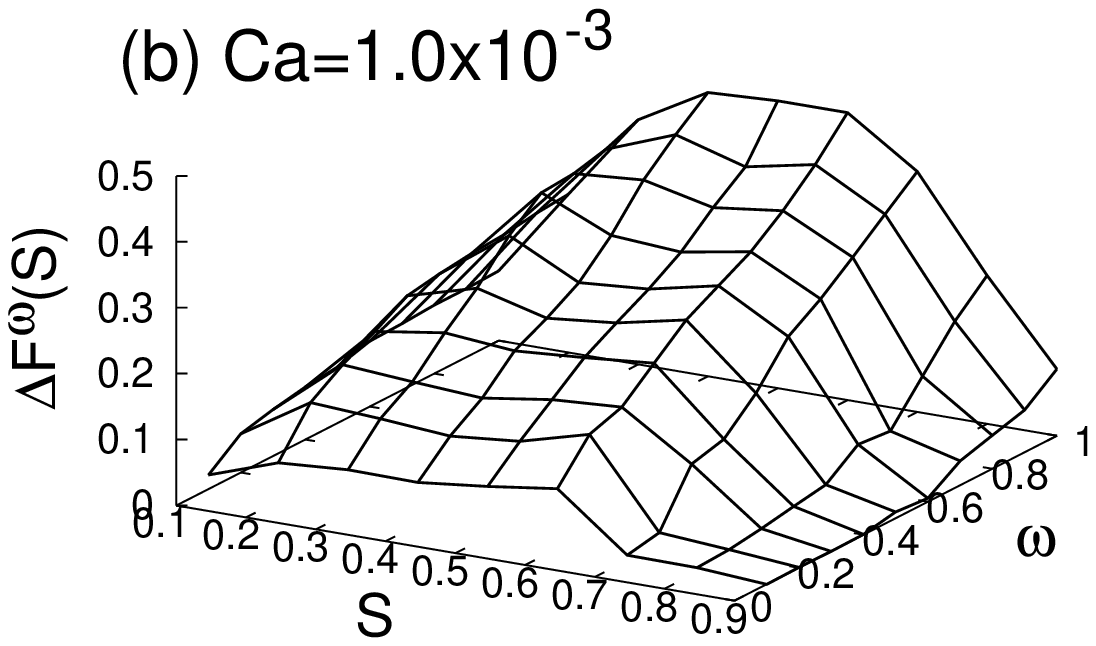,width=.22\textwidth}
\hfill}
\caption{\label{df-3d}Change in oil fractional flow $\Delta
  F^\omega(S)$ with saturation and system wettability. For the lower
  capillary number, a finite $\omega$ is required to have a non-zero
  value of $\Delta F^\omega(S)$ at high saturations.}
\end{figure}

As we consider only the case of viscosity ratio $M=1$, a symmetric
relationship between the fractional flow in pure water-wet and pure
oil-wet systems is expected. From straight-forward symmetry argument
we write the relation between the oil fractional flow in the two
different systems as,
\begin{equation}
F^1(S) = 1 - F^0(1-S)\;.
\label{f0}
\end{equation}
In Fig.\ \ref{flow-s}, $1 - F^0(1-S)$ is plotted by the dotted lines
and it matches nicely with $F^1(S)$ curve.

The wettability of real oil reservoirs generally have fractional
wetting conditions. We therefore measure $F^\omega(S)$ in different
systems with fractional values of $\omega$ from $0.1$ to $1.0$ in
intervals of $0.1$. The difference of $F^\omega(S)$ from that in pure
water-wet system, $\Delta F^\omega(S) = F^0(S) - F^\omega(S)$, is then
calculated. In Fig.\ \ref{df-3d}, $\Delta F^\omega(S)$ is plotted as a
function of saturation and wettability for the two different capillary
numbers. It can be seen that the surface is more flat for the higher
capillary number. Moreover, for the lower Ca there is a finite region
with $\Delta F^\omega(S)=0$ to be noticed. It is already observed in
Fig.\ \ref{flow-t}(b) that for high saturation a finite wettability
alteration $\omega_{tr}$ is required to permanently change the system
from a single-phase dominated regime.  For $\omega < \omega_{tr}$ we
have $\Delta F^\omega(S)=0$.

\begin{figure}[t]
\centerline{\hfill
  \psfig{file=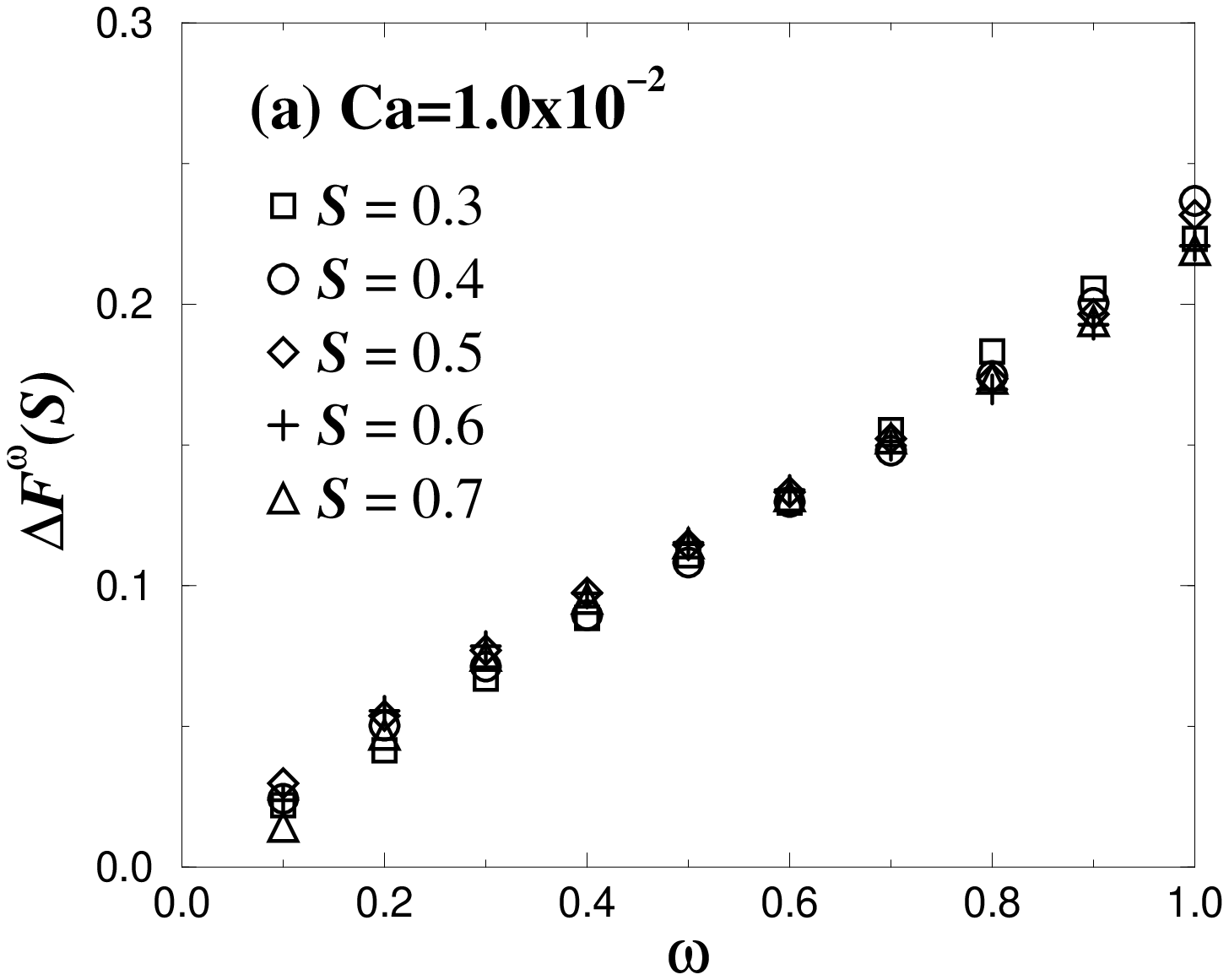,width=.22\textwidth}
  \hfill
  \psfig{file=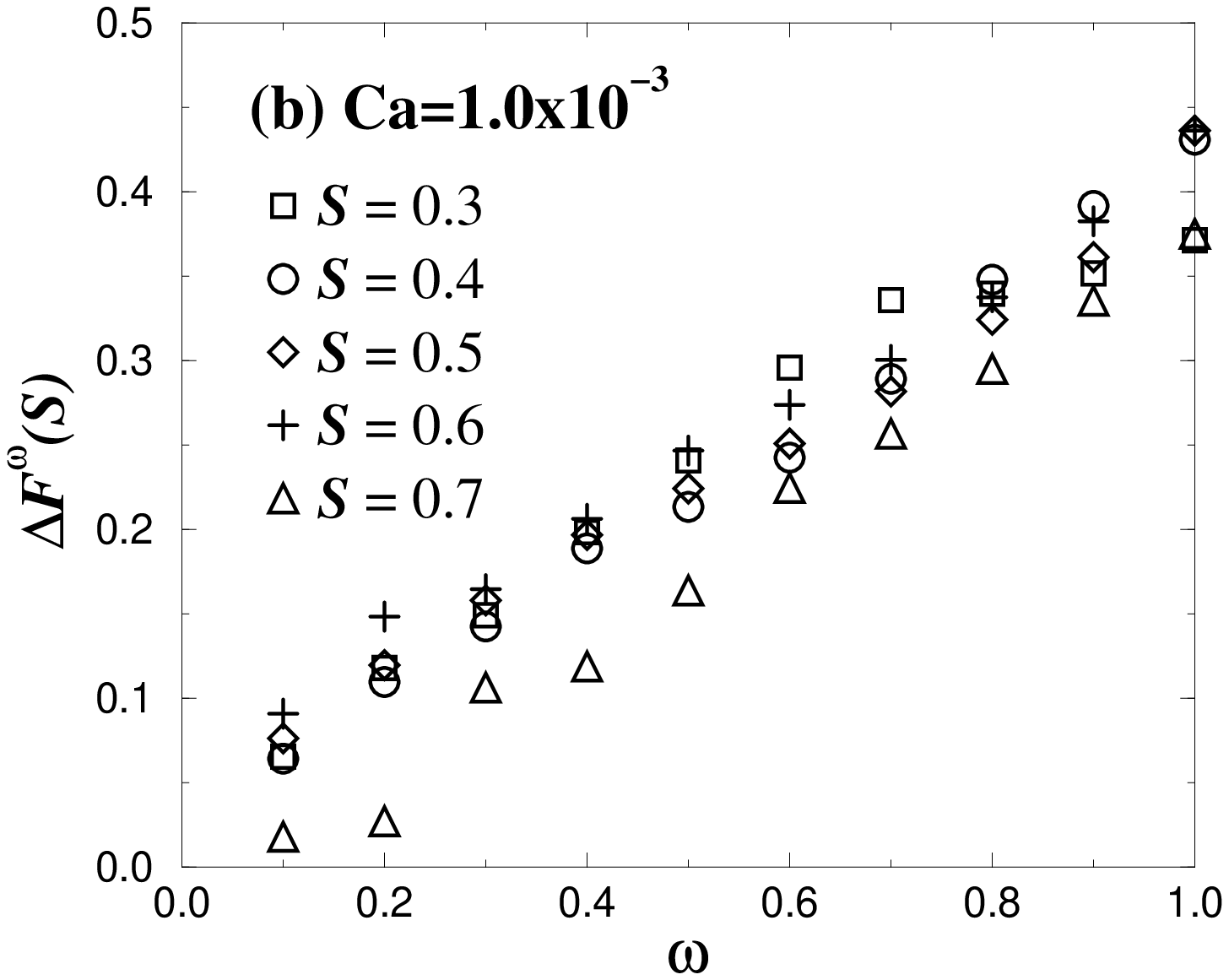,width=.22\textwidth}
  \hfill}
\caption{\label{df-slope}Change in the oil fractional flow $\Delta
  F^\omega(S)$ as a function of $\omega$ at intermediate range of
  saturation, from $S=0.3$ to $0.7$. It can be seen that $\Delta
  F^\omega(S)$ varies linearly with $\omega$.}
\end{figure}

In order to find the functional dependence of $\Delta F^\omega(S)$ on
$\omega$ in the regime where both the phases are mobile, $\Delta
F^\omega(S)$ is plotted against $\omega$ in Fig.\ \ref{df-slope} for
saturations from $S=0.3$ to $0.7$. The variation of $\Delta
F^\omega(S)$ with $\omega$ is found to be linear here with some
fluctuations. For $\omega=0$, $\Delta F^0(S)=0$. For $\omega=1$,
$\Delta F^1(S)$ directly follows from Fig.\ \ref{flow-s} as the
difference between the $F^0(S)$ and $F^1(S)$ curves. Assuming
linearity in between, one can write
\begin{equation}
\Delta F^\omega(S) = \omega \Delta F^1(S) = \omega\big[F^0(S) - F^1(S)\big],
\label{fw}
\end{equation}
which we can combine with eq.\ \ref{f0} to obtain
\begin{equation}
F^\omega(S) = F^0(S) - \omega\big[F^0(S) + F^0(1-S) - 1\big].
\end{equation}
This gives the functional dependence of fractional flow on both
saturation and system wettability. It connects fractional flow in a
{\it fractionally} wet system with that of a {\it purely} water-wet
system by a linear relation with $\omega$ and $S$.  As fractional flow
curves may be more easily obtained in samples of pure wetting
properties, this equation may be useful to have an estimate of
fractional flow when wettability is altered by, e.g., low-salinity
water injection.

As $\Delta F^1(S)$ is never negative it is clear that the fractional
flow of oil will never decrease with the increase of water-wetness of
the system. For a large viscosity contrast this result could vary, as
our symmetry argument requires viscosity matched fluids. It is
therefore requires further extensive numerical simulations to identify
the results in that regime.

As a conclusion, we study the effect as well as the detailed pore
level mechanism that causes the change in the steady state two-phase
flow properties due to wettability alteration in porous media. A
change in the system wettability causes a perturbation in the system's
flow pattern to destabilize any percolating and trapped immobile
clusters appeared in the steady state. In order to prevent forming
similar structures again, a sufficiently strong wettability alteration
is required depending upon the capillary number and saturation. When
both the phases are mobile, fractional flow of oil is found to
increase linearly with increasing water-wetness of the system. The
results of our simulation are general in nature and consistent with
the experimental results reported in literature qualitatively. They
show that the remobilization of fluid clusters is due to purely fluid
mechanical changes in the system induced by changes in the wetting
contact angle --- which in turn comes from the wettability alterations
of the porous medium.

The authors thank Glenn T{\o}r{\aa} for valuable discussions. The work
has been financed through Norwegian Research Council (NFR) Grant
No. 193298/S60.


\begin{thebibliography}{21}
\bibitem{r05} P. Roberts, {\it The end of oil: On the edge of a
  perilous new world\/} (Houghton Mifflin, New York, 2005).
\bibitem{a07} W. Abdallah {\em et al.}, Schlumberger Oilfield Review
  {\bf 19}, 44 (2007).
\bibitem{b98} M. J. Blunt, J. Pet. Sci. Eng. {\bf 20}, 117 (1998);
  J. S. Buckley, Curr. Opin. Coll. Int. Sci. {\bf 6}, 191 (2001);
  A. Skauge, K. Spildo, L. H{\o}iland and B. Vik,
  J. Pet. Sci. Eng. {\bf 57}, 321 (2007).
\bibitem{i11} J. Israelachivili, {\it Intermolecular and Surface
  Forces} (Academic Press, 3rd ed., 2011).
\bibitem{h91a} G. J. Hirasaki, in {\it Interfacial phenomena in
  petroleum recovery}, edited by N. R. Morrow (Marcell Dekker, New
  York, 1991), chs. 2 and 3.
\bibitem{k93} A. R. Kovscek, H. Wong, C. J. Radke, AIChe Journal, {\bf
  39}, 1072 (1993).
\bibitem{k98} R. Kaminsky, H. and C. J. Radke, Society of Petroleum
  Engineers, SPE 39087, 13 (1998).
\bibitem{tht99} M. T. Tweheyo, T. Holt and O. Tors\ae ter,
  J. Pet. Sci. Eng. {\bf 24}, 179 (1999).
\bibitem{tm97} G. Q. Tang and N. R. Morrow, SPE Res. Eng. {\bf 12},
  269 (1997); J. Pet. Sci. Eng.  {\bf 24}, 99 (1999).
\bibitem{sa00} D. C. Standnes and T. Austad, J. Pet. Sci. Eng.  {\bf
  28}, 123 (2000).
\bibitem{stk06} J. M. Schembre, G. Q. Tang and A. R. Kovscek,
  J. Pet. Sci. Eng. {\bf 52}, 131 (2006).
\bibitem{btm89} J. S. Buckley, K. Takamura and N. R. Morrow, SPE
  Res. Eng. {\bf 4}, 332 (1989).
\bibitem{amhb98} E. Aker, K. J. M\r{a}l\o y, A. Hansen and
  G. G. Batrouni, Transp. Porous Media {\bf 32}, 163 (1998); E. Aker,
  K. J. M\r{a}l\o y and A. Hansen, Phys. Rev. E {\bf 58}, 2217 (1998).
\bibitem{kah02} H. A. Knudsen, E. Aker and A. Hansen, Transp. Porous
  Media {\bf 47}, 99 (2002).
\bibitem{rh06} T. Ramstad and A. Hansen, Phys. Rev. E {\bf 73}, 026306
  (2006).
\bibitem{d92} F. A. L. Dullien, {\em Porous Media: Fluid Transport and
  Pore Structure} (Academic Press, San Diego, 1992).
\bibitem{w21} E. W. Washburn, Phys. Rev. {\bf 17}, 273 (1921).
\bibitem{bh88} G. G. Batrouni and A. Hansen, J. Stat. Phys. {\bf 52},
  747 (1988).
\bibitem{ss06} S. Sinha and S. B. Santra, Int. J. Mod. Phys. C {\bf
  17}, 1285 (2006).
\bibitem{kah02-2} H. A. Knudsen and A. Hansen, Phys. Rev. E {\bf 65},
  056310 (2002).
\bibitem{kh04} H. A. Knudsen and A. Hansen, Eurphys. Lett. {\bf 65},
  200 (2004).
\end{thebibliography}
\end{document}